\documentclass[12pt]{article}
\usepackage{times}
\usepackage{cite}
\usepackage{url}
\usepackage{hyperref}
\usepackage{graphicx}
\begin{document}
\title{Evaluating Multi-Instance DNN Inferencing on Multiple Accelerators of an Edge Device}
\author{Mumuksh Tayal\thanks{Cyber Physical Systems, Indian Institute of Science, Bangalore, INDIA. Email: mumukshtayal@iisc.ac.in} \and Yogesh Simmhan\thanks{Department of Computational and Data Sciences, Indian Institute of Science, Bangalore, INDIA. Email: simmhan@iisc.ac.in}}
\date{}
\maketitle

\begin{abstract}
Edge devices like Nvidia Jetson platforms now offer several on-board accelerators---including GPU CUDA cores, Tensor Cores, and Deep Learning Accelerators (DLA)---which can be concurrently exploited to boost deep neural network (DNN) inferencing. In this paper, we extend previous work by evaluating the performance impacts of running multiple instances of the ResNet50 model concurrently across these heterogeneous components. We detail the effects of varying batch sizes and hardware combinations on throughput and latency. Our expanded analysis highlights not only the benefits of combining CUDA and Tensor Cores, but also the performance degradation from resource contention when integrating DLAs. These findings, together with insights on precision constraints and workload allocation challenges, motivate further exploration of intelligent scheduling mechanisms to optimize resource utilization on edge platforms.

\end{abstract}

\section{Introduction and Motivation}
The widespread use of Deep Neural Network (DNN) models in applications such as smart city surveillance~\cite{8704334} and safe robot navigation 
\cite{learning_cbf}, 
and healthcare 
\cite{10.3389} 
has increased the demand for accelerated edge devices, such as Nvidia Jetson and Google Coral, for real-time decision-making. Such edge devices consume data from multiple sensor streams at high rates for efficient inferencing. So, it is critical to maximize their performance to increase the inferencing throughput and lower latency.

Among these, Nvidia's Jetson edge accelerators have become popular for industrial and IoT applications due to their small physical and energy footprint and high computing power that approach GPU workstations.  
E.g., the latest generation of Jetson Orin AGX used in this study
has a 12-core ARM-based CPU, an Ampere GPU with 2048 CUDA cores and 64 Tensor cores, two Deep Learning Accelerators (DLAs), and 32GB of shared memory, while consuming a peak power of 60W.

Existing studies compare the performance of various edge devices, with a focus on CPUs, CUDA cores and TPUs~\cite{sk2022characterizing,10.1145/3546192}. However, they either consider these co-located hardware resources independently or lack a comprehensive analysis of newer accelerators such as Tensor Cores and DLAs.
This poster addresses this gap by \textit{concurrently utilizing multiple on-board computing resources} for a Jetson AGX Orin edge device, such as CUDA cores, Tensor Cores and DLA, to evaluate the configurations that maximize the cumulative inferencing performance.
In particular, we run multiple instances of the same DNN model on the various hardware components, and
study the effects of different hardware combinations, DNN instance concurrency and batch sizes ($bs$) on throughput and latency. We also account for the precision constraints of each component, such as Tensor Cores' mixed precision and DLA's FP16 requirements, providing insights into the capabilities and limitations of edge devices that are less explored.



\begin{figure*}[t!]
\vspace{-0.1in}
\centering
    \includegraphics[width=1.2\textwidth]{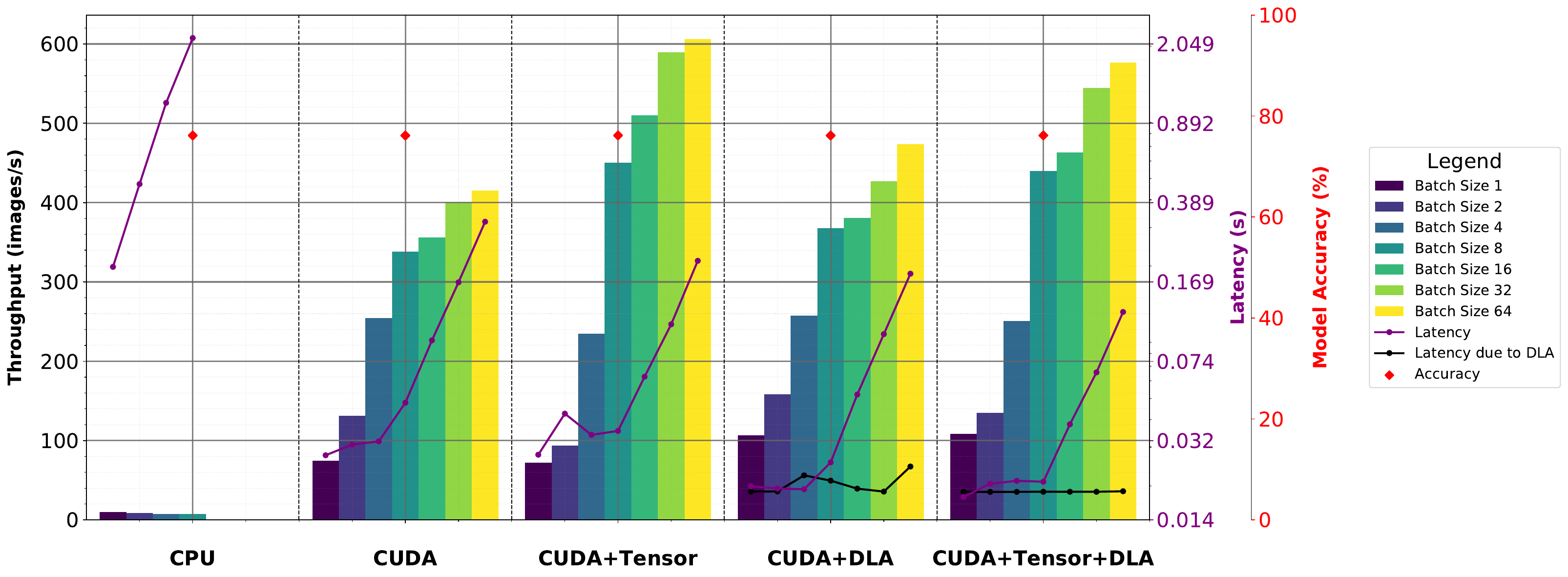}
    \caption{Throughput ($images/s$, left Y axis) and Response Latency ($s$, inner right Y axis) when concurrently using multiple hardware combinations (outer X axis), for different batch sizes ($bs$, inner X axis). Accuracy achieved (outer right Y axis) is similar in all the cases at $\approx 76\%$.}
    
    \label{fig:final_plot}
    \vspace{-0.2in}
\end{figure*}

\section{Experimental Setup}


The experiments are conducted on an Nvidia Jetson Orin AGX, utilizing its GPU CUDA cores, Tensor Cores and DLAs. 
We employ the a widely-used ResNet50 image classification CNN to benchmark the AGX Orin. We use the ImageNet1K-V1 dataset and images from the validation image subset as the inferencing samples.
PyTorch is used for deploying the model on CUDA cores and Tensor cores. Since PyTorch does not directly support DLAs, the model is converted to ONNX format and deployed using TensorRT when running on DLAs. 
Continuous streams of inference requests are generated such that each concurrently running instance receives a batch of images containing \textit{bs} number of images
to ensure saturation of the target component.

DLA requires separate compilation using TensorRT and independent execution. Here, we run a DNN instance independently on the DLA to consume the inference requests, and another DNN instance on the other hardware resources for that experiment, separately consuming the requests. In contrast, the Nvidia compiler handles the workload distribution for a single DNN instance among CUDA and Tensor Cores used together.
To ensure consistency, we run 2 instances of the DNN in the other cases as well, e.g., 2 DNN inferencing processes on the CPU or on the CUDA Cores.

We benchmark the performance under three scenarios:
\begin{enumerate}
    \item Running multiple instances of the CNN model on individual components to measure their peak performance.
    \item Running multiple instances on all components simultaneously to assess overall throughput.
    \item Running pairs of components to evaluate efficiency and resource sharing, monitored via the \textit{jtop} module.
\end{enumerate}


CUDA cores operate under full \textit{FP32} precision, while DLA uses \textit{FP16} and Tensor cores use Automatic Mixed Precision (AMP).
Due to architectural constraints, we cannot independently run the ResNet model on Tensor Cores and DLA to isolate their performance. Since Tensor Cores use AMP, they still rely on CUDA cores selectively,
while certain layers of ResNet50, such as GlobalAveragePool, are incompatible with DLA, leading them to a fallback to the CUDA cores. So we run the model on these combinations: CPU only \textit{(baseline)}, CUDA cores only \textit{(baseline)}, CUDA+Tensor Cores, CUDA Cores+DLA, and CUDA+Tensor Cores+DLA.
As we report, the CPU gives minimal inference throughput, and is excluded from the other combinations.

\section{Results}
%
Despite using different precisions for the accelerators,
the prediction accuracy 
for all hardware combinations remains largely unaffected, at $\approx 76\%$ for all 50k images in the validation dataset (Fig.~\ref{fig:final_plot}).
So we focus on the throughput and latency performance in the ensuing discussion.

\textit{Single Component Utilization Evaluation.~}
We measure the throughput of CPU and CUDA cores to establish the baseline performance. GPU CUDA cores achieve the best throughput of $415.2$~images/s at $bs=64$, while CPU cores have the highest throughput of $10.1$~images/s with just $bs=1$ (Fig.~\ref{fig:final_plot}). As expected, the GPU outperforms the CPU by $\approx 50\times$, primarily due to the higher parallelism exploited by the CUDA cores for
larger batch sizes. The CPU cores are unable to realize comparable benefits.

\textit{Pairwise Component Utilization Evaluation.~}
The evaluation of combined components revealed distinct throughput advantages across batch sizes. The integration of Tensor Cores with CUDA Cores, employing AMP, showed superior throughput at larger batch sizes compared to CUDA Cores alone. 
In contrast, the DLA and CUDA Core pairing yielded a notable throughput improvement of $40$--$50$~images/s at smaller batch sizes, considering that DLA operates at a lower precision of FP16. But the model accuracy is still retained. 
In all scenarios involving DLA, two separate processes run concurrently, necessitating that their latencies be reported independently (Fig.~\ref{fig:final_plot}). Although the Tensor Core and CUDA Core combination enhanced performance with increasing batch sizes, the throughput for the DLA and CUDA configuration showed diminishing returns at larger batch sizes due to contention for CUDA Core resources.

\textit{Multi-Component Utilization Evaluation.~}
When all three components operated simultaneously, the system exhibited suboptimal performance due to resource contention, which we are examining. Throughput consistently increased with larger batch sizes; however, when comparing the results of the CUDA and Tensor Cores combination -- known to achieve maximum throughput -- the throughput of the CUDA+Tensor+DLA configuration was lower at higher batch sizes, e.g., $29.5$~images/sec fewer for $bs=64$.


\section{Conclusion and Future Work}
Our results show the throughput benefits of concurrent execution of workloads using multiple accelerators on edge devices. But, it also highlights the challenges of resource contention that degrades performance when all three components, DLA, CUDA and Tensor Cores, are used.
As future work, we will examine strategic load allocation and configuration for heterogeneous components based on their characteristics, and to adapt to the varying workload requirements. 




\label{References}
\bibliographystyle{IEEEtran}
\bibliography{references.bib}

\end{document}